\documentclass[12pt,a4paper,dvips]{article}
\usepackage{epsfig,wrapfig,times,mathptm} 
\renewcommand{\section}[1]{\vspace{6pt} \noindent\mbox{#1} \newline \noindent}
\renewcommand{\subsection}[1]{\vspace{6pt} \noindent\mbox{\underline{#1}} 
\newline \noindent}
\renewcommand{\subsubsection}[1]{\vspace{6pt} \noindent\mbox{\underline{#1}}
\noindent}

\newfont{\sansb}{cmssbx10}
\newfont{\sans}{cmss10}

\begin{document}
{\small HE 1.2.36 \vspace{-24pt}\\}     
{\center \LARGE ABSENCE OF GREISEN-ZATSEPIN-KUZMIN CUTOFF 
AND STABILITY OF UNSTABLE
PARTICLES AT VERY HIGH ENERGY, AS A CONSEQUENCE OF LORENTZ SYMMETRY VIOLATION
\vspace{6pt}\\}
L. Gonzalez-Mestres$^{1,2}$ \vspace{6pt}\\
{\it $^1$Laboratoire de Physique Corpusculaire, Coll\`ege de France, 
75231 Paris Cedex 05 , France\\
$^2$Laboratoire d'Annecy-le-Vieux de Physique des Particules, 74941 
Annecy-le-Vieux Cedex,
France
 \vspace{-12pt}\\}
{\center ABSTRACT\\}
Special relativity has been tested at low energy with great accuracy, but
its extrapolation to very high-energy phenomena is much less well established.
Introducing a critical distance scale, $a$ , 
below $10^{-25}~ cm$ (the wavelength
scale of the highest-energy observed cosmic rays) allows to consider mo- dels,
compatible with standard tests of special relativity, where a small
violation of Lorentz symmetry ($a$ can, for instance, be the Planck length)
produces dramatic effects on the properties of high-energy cosmic rays.
Not only the Greisen-Zatsepin-Kuzmin (GZK) 
cutoff on very high-energy protons and
nuclei does no longer apply, but particles which are unstable at low energy
(neutron, several nuclei, some hadronic resonances 
like the $\Delta ^{++}$...)
would become stable at very high energy. The muon would also become stable
or very long lived at very high energy if one of the two neutrinos 
associated to the light
charged leptons (electron, muon) has a mass. Similar considerations apply to 
the $\tau $ lepton.
We discuss several possible
scenarios originating these phenomena, as well as the cosmic ray energy
range (well below the energy scale associated to the fundamental length)
and experiments where they could be detected. Observable effects
are predicted for the highest-energy cosmic rays.

\setlength{\parindent}{1cm}
\section{LORENTZ SYMMETRY AS A LOW-ENERGY LIMIT}
Low-energy experiments (Lamoreaux, Jacobs, Heckel, Raab and Forston, 
1986 ; 
Hills and Hall, 1990) confirm Lorentz invariance to an impressive accuracy.
However, the extrapolation between these results and high-energy phenomena
is far from obvious. Figures can change by 34 orders of magnitude between
$keV$ and $10^{20}~eV$ scales 
if Lorentz symmetry violation is proportional to $(k~a)^2$ where $k$ is the
wave vector scale and $a$ a fundamental length. Such a behaviour seems to
be characteristic of many models where local Lorentz invariance is broken
through non-local phenomena at the fundamental length scale (f.i. the Planck 
scale). These models lead to a dispersion relation of the form
(Gonzalez-Mestres, 1997):
\equation
E~~=~~(2\pi )^{-1}~h~c~a^{-1}~e~(k~a)
\endequation
where $E$ is the energy of the particle,
$h$ the Planck constant, $c$ the speed of light, 
$a$ a fundamental length scale that we can
naturally identify with the Planck length (but other choices of the
fundamental distance scale are possible), $k$
the wave vector modulus and 
$[e~(k~a)]^2$ is a convex        
function of $(k~a)^2$ obtained from vacuum dynamics.  
We have checked that this is also a fundamental property of old 
scenarios breaking local Lorentz invariance (f.i. R\'edei, 1967), 
although such a
phenomenon seems not to have been noticed by the authors. For a particle of 
mass $m$ ,
an ansatz based on an isotropic, continuous modification of the Bravais 
lattice dynamics is (Gonzalez-Mestres, 1997):
\equation
e~(k~a)~~=~~[4~sin^2~(ka/2)~+~(2\pi ~a)^2~h^{-2}~m^2~c^2]^{1/2}
\endequation
whereas we have found that simple extensions of the ansatz by R\'edei (1967) 
lead to expressions like:
\equation
e~(k~a)~~=~~[10~+~30~(k~a)^{-2}~cos~(k~a)~-~30~(k~a)^{-3}~sin~(k~a)
~+~(2\pi ~a)^2~h^{-2}~m^2~c^2]^{1/2}
\endequation
which has similar properties to (2). 
In both cases, and rather generally, we find
that, at wave vector scales below the inverse of the fundamental length scale,
Lorentz symmetry violation in relativistic kinematics can be parameterized
writing:
\equation
e~(k~a)~~\simeq ~~[(k~a)^2~-~\alpha ~(k~a)^4~+~(2\pi ~a)^2~h^{-2}~m^2~c^2]^{1/2}
\endequation
where $\alpha $ is a positive 
constant between $10^{-1}$ and $10^{-2}$ . At high energy, we can write:
\equation
e~(k~a)~~\simeq ~~k~a~[1~-~\alpha ~(k~a)^2/2]~
+~2~\pi ^2~h^{-2}~k^{-1}~a~m^2~c^2
\endequation
and, in any case, we expect observable kinematical effects when the term 
$\alpha (ka)^3/2$ becomes as large as the term 
$2~\pi ^2~h^{-2}~k^{-1}~a~m^2~c^2$ .
This happens at:
\equation
E~~\simeq (2\pi )^{-1}~h~c~k~~\approx ~~ \alpha ^{-1/4}~
(h~c~a^{-1}/2\pi)^{1/2}~(m~c^2)^{1/2}
\endequation
Thus, contrary to conventional estimates  
of local Lorentz symmetry breaking predictions (Anchordoqui, Dova, G\'omez Dumm
and Lacentre, 1997)
where the modification of 
relativistic kinematics is ignored, 
observable effects will be produced at wavelength scales well above 
the critical length.  
For a proton or a neutron, and taking $a~\approx ~10^{-33}~cm$ ,
this corresponds to $E~\approx 10^{19}~eV$ , an energy scale below the highest
energies at which cosmic rays have been observed. Similar considerations apply
to nuclei and would apply to muons, pions 
and $\tau $ leptons if these particles were stable.
It must be realized that, for a proton 
at $E~\approx ~ 10^{20}~eV$ and with the above value of $a$ , one would have:
\equation
\alpha ~(k~a)^2/2~~\approx ~~10^{-18} ~~\gg ~~2~\pi ^2~h^{-2}~k^{-2}~m^2~c^2~~
\approx 10^{-22}
\endequation
so that, although $\alpha (ka)^3/2$ is indeed very small as compared to the
value of $e~(k~a)$ , the term $2~\pi ^2~h^{-2}~k^{-1}~a~m^2~c^2$ represents
an even smaller fraction of this quantity. We therefore expect corrections
to relativistic kinematics to play a crucial role at the highest 
cosmic ray energies.
Although Lorentz symmetry certainly reflects to a very good approximation
the reality of physics at large distance scales and can therefore be 
considered as the low-energy limit of particle kinematics, no existing 
experimental result proves that it applies with the same accuracy to 
high-energy cosmic rays. In view of the above considerations, 
the question deserves serious practical study 
in close connexion with high-energy experiments. In what follows, we discuss
two expected consequences of local Lorentz symmetry violation, assuming the
value of $c$ in (1) to be a universal constant.

\section{THE GZK CUTOFF DOES NO LONGER APPLY}
A proton with $E~>~10^{20}~eV$ interacting with a cosmic
microwave background photon would be sensitive to the above corrections
to relativistic kinematics.
For instance,
after having absorbed a $10^{-3}~ eV$ photon moving in the opposite direction,
the proton gets an extra $10^{-3}~ eV$ energy, whereas its momentum is
lowered by $10^{-3}~ eV/c$ . In the conventional scenario with exact Lorentz
invariance, this is enough to allow the excited proton to decay into a
proton or a neutron
plus a pion, losing an important part of its energy. However, it
can be checked (Gonzalez-Mestres, 1997) 
that in our scenario with Lorentz invariance violation such
a reaction is strictly forbidden.  Elastic $p~+~\gamma $ scattering 
is permitted, but allows the
proton to release only a small amount of its energy.
The outgoing photon energy for an incoming $10^{20}~eV$
proton cannot exceede
$\Delta E^{max} ~ \approx 10^{-5}~E~=~10^{15}~eV$ instead of the value
$\Delta E^{max} ~ \approx ~10^{19}~eV$ obtained with exact Lorentz invariance.
Similar or more stringent bounds exist
for channels involving lepton production. Furthermore, obvious phase
space limitations will also lower the event rate, as compared to standard
calculations using exact Lorentz invariance which predict photoproduction
of real pions at such cosmic proton energies. The effect seems strong enough
to invalidate the Greisen-Zatsepin-Kuzmin cutoff 
(Greisen, 1966; Zatsepin and Kuzmin, 1966) and explain the
existence of the highest-energy cosmic rays. It will become more
important at higher energies, as we get closer to the $a^{-1}$ wavelength
scale. Similar arguments apply to heavy nuclei, again invalidating the GZK
cutoff. Since, in both cases, the cosmic ray energy was expected to degrade
over distances $\approx 10^{24}~m$ 
according to conventional estimates, the correction by several orders
of magnitude we just introduced applies to distance scales much larger than the
estimated size of the presently observable
Universe. Obvioulsy, our result is limited by the history of the Universe, as
cosmic rays coming from distances closer and closer 
to $c^{-1}$ times the horizon
size will be older and older and, at early times, will have been confronted
to rather different scenarios. Nevertheless, the above modification of 
relativistic kinematics allows much older cosmic rays to reach earth nowadays. 

A previous attempt to explain the experimental absence of the predicted
GZK cutoff by 
Lorentz symmetry violation at high energy (Kirzhnits and Chechin, 1972)
led the authors to consider an expansion in powers of $\gamma ^4$ , where 
$\gamma ~=~(1~-~v^2c^{-2})^{-1/2}$ , $v$ is the speed of the particle
and the coefficient of the linear term
in $\gamma ^4$ had to be arbitrarily
tuned to $\approx 10^{-44}$ in order to produce
an effect of order 1 for a $10^{20}~eV$ proton (therefore leading to a
divergent expansion at higher energies). No such problems are encountered
in our approach, where the required orders of magnitude come out quite 
naturally. If the absence of GZK cutoff is indeed due to the kinematics defined 
(1), it allows in principle to set a lower bound on the value of the 
fundamental length (around $10^{-34}~cm$).

\section{UNSTABLE PARTICLES MAY BECOME STABLE AT VERY HIGH ENERGY}
In standard relativity, we can compute the lifetime of any unstable particle
in its rest frame and, with the help of a Lorentz transformation, obtain
the Lorentz-dilated lifetime for a particle moving at finite speed. 
The same procedure had been followed in previous estimates of the predictions
of local Lorentz symmetry breaking
(Anchordoqui, Dova, G\'omez Dumm
and Lacentre, 1997) for the decay of high-energy particles.
This is no longer possible with the kinematics defined by (1). 
Instead, two results are obtained (Gonzalez-Mestres, 1997):

i) Unstable particles with at least two massive particles in the final state
of all their decay channels become stable at very high energy,
as a consequence of Lorentz symmetry violation through (1). 
A typical order of magnitude for the energy $E^{st}$ 
at which such a phenomenon occurs 
is: 
\equation
E^{st}~~\approx ~~c^{3/2}~
h^{1/2}~(a~m_2)^{-1/2}~(m^2~-~m_1^2~-~m_2^2)^{1/2}
\endequation
where: a) $m$ is the mass of the decaying particle; b) 
we select the two 
heaviest particles of the final product of each decay channel, and
$m_2$ is the mass of the lightest particle
in this list; c) $m_1$ is the mass of the heaviest particle produced together
with that of mass $m_2$ . 
With $a~\approx ~10^{-33}~cm$ , {\bf the neutron
would become stable} for $E~\stackrel{>}{\sim }~10^{20}~eV$ .
At the same energies or slightly above, {\bf some unstable
nuclei would also become stable}.
Similarly, {\bf some hadronic resonances} (e.g. the $\Delta ^{++}$ , whose 
decay
product contains a proton and a positron)
{\bf would become stable} at
$E~\stackrel{>}{\sim }~10^{21}~eV$ .
Most of these objects will
decay before they can be accelerated to such energies, but
they may result of a collision at very high energy or of
the decay of a
superluminal particle (Gonzalez-Mestres, 
1996). 
The study of very high-energy cosmic rays can thus
reveal as stable particles objects which would be unstable if produced at
accelerators.
If one of the light neutrinos
($\nu _e$ , $\nu _{\mu }$) has a mass in the $\approx ~10~eV$
range, {\bf the muon would
become stable} at energies above $\approx 10^{22}~eV$ . Weak neutrino mixing
may restore muon decay, but with very long lifetime. {\bf 
Similar considerations 
apply to the $\tau $ lepton}, 
which would become stable above 
$E~\approx 10^{22}~eV$ if the mass of the $\nu _{\tau }$ is $\approx ~100~eV$
but, again, a decay with very long lifetime can be restored by 
neutrino oscillations. 

ii) In any case, unstable particles live longer than naively expected with exact
Lorentz invariance and, at high enough energy, 
the effect becomes much stronger than previously estimated
(Anchordoqui, Dova, G\'omez Dumm
and Lacentre, 1997)
ignoring the small violation of relativistic kinematics. 
At energies well below the stability region, partial decay rates are already 
modified by large factors leading to observable effects. 
Irrespectively of whether 
$m_2$ vanishes or not, the phenomenon
occurs at $E~\stackrel{>}{\sim }~c^{3/2}~h^{1/2}~(m^2~-~m_1^2)^{1/4}~a^{-1/2}$
($\approx 10^{18}~eV$ for $\pi ^+~\rightarrow ~e^+~+~\nu _e$ , if 
$a~\approx ~10^{-33}~cm$). The effect has a sudden, sharp rise, since a fourth 
power of the energy is involved in the calculation.

\section{CONCLUDING REMARKS}
For similar reasons, a small violation of the universality of $c$ would not
necessarily produce the Cherenkov effect in vacuum  
considered by Coleman and Glashow (1997) for high-energy cosmic rays. 
The mechanism we just described competes with those considered
in their discussion and tends to compensate their effect: 
therefore, the bounds obtained by these authors do not apply to our  
ansatz. On the other hand, 
the discussion of velocity oscillations of neutrinos
presented by Glashow, Halprin, Krastev, Leung and Pantaleone (1997) for the
low-energy region is compatible with our theory. However, the universality 
of $c$ seems natural in most unified 
field theories (whereas that of the mass is 
naturally violated) and preserves the Poincar\'e relativity principle 
(Poincar\'e, 1905) in the low-momentum limit. In any case, if Lorentz symmetry
is broken and an absolute rest frame exists, high-energy particles are
indeed different physical objects from low-energy particles.

It is also interesting to lower the value of $a^{-1}$ down to the 
wave vector of the highest-energy cosmic rays, $\approx 10^{25}~cm^{-1}$. 
Then, a stable neutron is 
predicted at energies $\stackrel{>}{\sim }~10^4~TeV$ and, with respect to the
above estimates for other particles, 
the energy threshold for stability is to be lowered by a
factor $\approx ~10^{-4}$ .
For similar reasons, the
departure from standard relativistic values for partial decay rates would 
start at $E~\stackrel{>}{\sim }~100~TeV$ for the 
$\pi ^+~\rightarrow ~e^+~+~\nu _e$ channel. Not only lifetimes do not follow
relativistic formulae, but partial branching ratios become energy-dependent
and are sensitive to the masses of the produced particles.
Data on high-energy cosmic
rays contain information relevant to these phenomena and should be carefully
analyzed. 
Cosmic rays seem to indeed be able 
to test the predictions of (1) 
and set upper bounds on the fundamental length $a$ .
Experiments like AUGER and AMANDA present great potentialities in this respect.
Very high-energy data may even provide a way to measure
neutrino masses and mixing, as well as other parameters related to phenomena
beyond the standard model.
Even if the energy is in principle
too low, lifetime measurements at LHC energies are also
worth
performing. Because of its stability at very high energy, the neutron becomes
a serious candidate to be a possible primary of the highest-energy cosmic 
ray events. 

\section{ACKNOWLEDGEMENTS}
It is a pleasure to thank J. Gabarro-Arpa, as well as 
colleagues at LPC Coll\`ege de France, for useful
discussions and remarks.

\section{REFERENCES}
\setlength{\parindent}{-5mm}
\begin{list}{}{\topsep 0pt \partopsep 0pt \itemsep 0pt \leftmargin 5mm
\parsep 0pt \itemindent -5mm}
\vspace{-15pt}
\item
Anchordoqui, L., Dova, M.T., G\'omez Dumm, D. and Lacentre, P., 
{\it Zeitschrift f\"{u}r Physik C} 73 , 465 (1997).
\item
Coleman, S. and Glashow, S.L., "Cosmic Ray and Neutrino Tests of Special 
Relativity", paper hep-ph/9703240 of LANL (Los Alamos) 
electronic archive (1997).
\item
Glashow, S.L., Halprin, A., Krastev, P.I., Leung, C.N. and Pantaleone, J.,
"Comments on Neutrino Tests of Special Relativity", paper hep-ph/9703454 
of LANL 
electronic archive (1997).
\item
Gonzalez-Mestres, L., "Physical and Cosmological Implications of a Possible
Class of Particles Able to Travel Faster than Light", contribution to the
28$^{th}$ International Conference on High-Energy Physics, Warsaw July 1996 .
Paper hep-ph/9610474 of LANL electronic archive (1996).
\item
Gonzalez-Mestres, L., "Vacuum Structure, Lorentz Symmetry and Superluminal
Particles", paper physics/9704017 of LANL  
electronic archive (1997).
\item
Greisen, K., {\it Phys. Rev. Lett.} 16 , 748 (1966).
\item
Hills, D. and Hall, J.L., {\it Phys. Rev. Lett.} 64 , 1697 (1990).  
\item
Kirzhnits, D.A., and Chechin, V.A., {\it Soviet Journal of Nuclear Physics}, 
15 , 585 (1972).
\item
Lamoreaux, S.K., Jacobs, J.P., Heckel, B.R., Raab, F.J. and Forston, E.N., 
{\it Phys. Rev. Lett.} 57 , 3125 (1986).
\item
Poincar\'e, H., Speech at the St. Louis International Exposition of 1904 ,
{\it The Monist 15} , 1 (1905).
\item
R\'edei, L.B., {\it Phys. Rev.} 162 , 1299 (1967).
\item
Zatsepin, G.T. and 
Kuzmin, V.A., {\it Pisma Zh. Eksp. Teor. Fiz.} 4 , 114 (1966).

\end{list}

\end{document}